\documentclass[sigconf]{acmart}

\usepackage{booktabs} 

\setcopyright{rightsretained}

\acmDOI{10.475/123_4}

\acmISBN{123-4567-24-567/08/06}

\acmConference[WI-IAT 2017]{IEEE/WIC/ACM International Conference on Web Intelligence 2017}{August 2017}{Leipzig, Germany} 

\begin{document}
\title{Emotions make cities live. Towards mapping emotions of older adults on urban space.}


\author{Radoslaw Nielek}
\orcid{0000-0002-5794-7532}
\affiliation{%
  \institution{Polish-Japanese Academy of Information Technology}
  \streetaddress{ul. Koszykowa 86}
  \city{02-008 Warsaw} 
  \state{Poland} 
}
\email{nielek@pja.edu.pl}

\author{Miroslaw Ciastek}
\affiliation{%
  \institution{Polish-Japanese Academy of Information Technology}
  \streetaddress{ul. Koszykowa 86}
  \city{02-008 Warsaw} 
  \state{Poland} 
}
\email{miroslaw.ciastek@pja.edu.pl}

\author{Wies\l{}aw Kope\'{c}}
\affiliation{%
  \institution{Polish-Japanese Academy of Information Technology}
  \streetaddress{ul. Koszykowa 86}
  \city{02-008 Warsaw} 
  \state{Poland} 
}
\email{kopec@pja.edu.pl}

\renewcommand{\shortauthors}{R. Nielek et al.}

\begin{abstract}
Understanding of interaction between people and urban spaces is crucial for inclusive decision making process. Smartphones and social media can be a rich source of behavioral and declarative data about urban space, but it threatens to exclude voice of older adults. The platform proposed in the paper attempts to address this issue. A universal tagging mechanism based on the Pluchik Wheel of Emotion is proposed. Usability of the platform was tested and prospect studies are proposed.

\end{abstract}

%
%



\maketitle

\section{Introduction}

Urban space without people and emotions associated with it resembles more a scenography than a city. The concept psychogeography, connecting psychological factors with space, was coined in early fifties by Ivan Chtcheglov \cite{chtcheglov1953formulary} and followed by progressing inclusion of citizens into urban planning and decision process.

A variety of methods helping citizens to participate in local initiatives have been developed -- i.a.: referendums, participatory budgeting and deliberative polling. Ubiquitous smartphones, cameras and sensors allowed gathering not only declarative but also behavioral data. People can vote and decide \textit{"by walking"} to particular places or posting pictures in social media instead of participating in polls. Guthier et al. sketched the idea of affect-aware city \cite{guthier2015affect}. This is a positive phenomenon that improves our understanding of the city dynamics but also carries risks. Next to the privacy issues, there is also a chance to miss opinion of older adults who rarely use social media extensively. It is worth noticing that the main problem is less and less related, at least for developed countries, with the lack of either required skills or devices and more with feeling that social media is not the place for the elderly.

Therefore a dedicated platform has been developed to strength the participation of the elderly in the process of taking community decisions. The platform mixes two modes: implicit and explicit one. Instead of directly asking about some issues, users can tag either pictures pushed by researchers (polling-like approach) or made by themselves (behavioral data approach). Tagging is done by selecting one of eight basic emotions from a wheel-like diagram (fig. \ref{fig:plutchik}) proposed by Robert Plutchik \cite{plutchik1980emotion} as it was proved that the understanding of emotions by older adults does not degenerate with age \cite{phillips2002age}. Pictures made by users have also a geo-tagging information attached. Mobile application was developed for tables (Android 4.4 or higher) and designed to be easy to use by older adults. Structured tagging approach has been selected as it is usually easier to do and has chance to deliver higher quality results \cite{bar2008structured}.

The remaining part of the paper is organized as follows. Related works are presented in the second section. The following section contains a description of the platform. Fourth section sketches preliminary usability tests conducted during developing phase. The last section summarizes the outcomes and describes ongoing projects related to prospect application of the platform.

\section{Related works}

\subsection{Emotion tagging}
\begin{figure}
\centering
\includegraphics[height=130px]{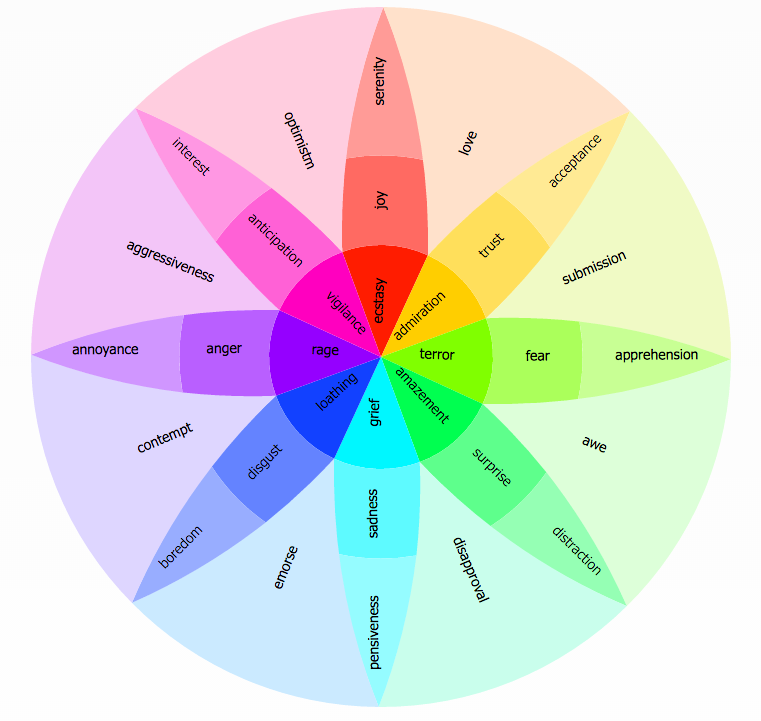}
\caption{Plutchik wheel of emotion.}
\label{fig:plutchik}
\end{figure}

Emotions are indispensably connected with peoples' lives, but at the same time they are extremely hard to quantify and describe. One of the oldest, but also the most well-known, attempt to put emotions in order was done by Robert Plutchik \cite{plutchik1980emotion}. Plutchik's Wheel of Emotions (PWE) depicted in Fig. \ref{fig:plutchik} is composed of eight bipolar emotions placed on the wheel. Although initially PWE was not meant for tagging emotions, researchers successfully used it for that purpose \cite{DBLP:journals/corr/TrompP14,mohammad2013crowdsourcing}. Runge et al. adopted Plutchik wheel of emotion to special requirements of mobile touchscreen devices \cite{runge2016tag}. Another well-established psychological tool for tagging emotion is Geneva Emotion Wheel (third version is the most up-do-date) \cite{schlegel2015introducing}. The most recent version (GEW v.3) contains 20 emotions (all with two labels) and five-point scale for each. GEW in comparison to PWE is much more detailed but at the same time more complex to understand and difficult to navigate on a relatively small touchscreen.

\subsection{Sensing urban environment}
\label{sec:sensing}
Mobile phones and user generated content are limitless sources of information about people's behaviors. So, it is not surprising that these data stream was extensively used for studying interactions between people and urban space. Researchers showed that the mobile phone can be aware of emotion state of the user \cite{Nielek:2010:EAM:1948645.1948662} and mobile phone usage patterns can help to predict negative emotion \cite{hung2016predicting}. Guthier at el. developed a system for extracting emotions (i.e. pleasantness, arousal, dominance and upredictability) from geo-tagged Twitter posts and visualize it on the map \cite{Guthier:2014:DVE:2661704.2661708}.

Next to the extraction of emotions from texts, researchers also tried to use pictures. Quercia et al. \cite{Quercia:2014:ACM:2531602.2531613} showed people two pictures from London and asked them to compare them on three dimension: beautifulness, quietness and happiness. Next, machine learning algorithms were used to identify in pictures objects influencing people's feeling. Collected data were also used for drawing a "happy map" of the city \cite{Quercia:2015:CHS:2740908.2741717} and recommending the most beautiful and happy paths between two points \cite{Quercia:2014:SPH:2631775.2631799}. Santani et al. applied deep learning algorithm (convolutional neural network) to automatically infer place ambiance \cite{Santani:2016:ILP:2964284.2967261}.

Similar approach was applied by Ruiz-Correa et al. to investigate impressions of outdoor urban spaces by young (16-18 years old) inhabitants of a city in Central Mexico \cite{Ruiz-Correa:2014:YCC:2694768.2694785}. Standard scales in social sciences were applied to evaluate danger, accessibility and dirtiness. By analyzing over 5000 ratings of images showing built environment and people inhabiting it, Traunmueller et al. revealed that familiarity is the most importnat predictor of the sense of safety \cite{Traunmueller:2016:YSE:2962735.2962761}.

Oliveira and Painho \cite{de2015emotion} proposed an Ambient Geographic Information approach extending a well-known idea of Geographic Information Systems (GIS) with layer dedicated for people's perception and feelings extracted from Twitter, Flickr, Instagram and Facebook. Based only on data from Foursquare and OpenStreetMap, Venerandi et al. tried to measure urban deprivation \cite{Venerandi:2015:MUD:2675133.2675233}.

Surprisingly, limited studies have been conducted on collecting opinions of older adults (65+) about their environment through ICT (and mobile applications in particular). The notable exception is work published by Thome et al. \cite{thome2014mobile}. Researchers from the University of New South Wales prepared a platform for collecting opinions of older adults about liveability in the city. The platform is composed of mobile application (designed for iPad) and server scripts that takes care of storing and visualizing opinions. 

The concept of places and pictures associated with emotions seems to be one of the easiest to understand for older adults even with some cognitive disorders \cite{capstick2015place}.

\section{Platform design}

As it is shown in section \ref{sec:sensing}, collection of pictures tagged by citizens might be a very powerful tool for reasoning about social dimensions of urban space. At the same time, solutions dedicated to older adults have to take into account their lower proficiency in the use of information technology, mild cognitive disorders and sight problems. Entire span of different possible tasks and questions that might arise in the participatory budgeting and urban planning process should be mapped on a very simple interface. Therefore, the platform is build around simple, atomic task of tagging picture with one emotion by placing it on the PWE. These tasks can be grouped into an experiment\footnote{A term \textit{"experiment"} is used in the paper but it might be also \textit{"poll"} or \textit{"referendum"}. For policy makers single experiment might be understand as an single issue (or a set of interconnected issues) be resolved or decided.}. 

There are two basic modes of the experiment that can be selected during the configuration phase: evaluating pictures prepared by researcher or evaluating pictures made by user (pictures are uploaded to the server and available to the researcher along with evaluations and location). For the first mode, a researcher has to select and upload pictures and decide about the order of displaying pictures to the participants (random order for each participant is also possible). Next, invitations to participate have to be distributed. It can be done either by sending email with an appropriate link or by using a QR Code (printed or displayed).

The platform supports multiple languages and select the appropriate one based on the local settings of used device. As the PWE is an established standard it is available in many different languages with names of emotions already translated and validated. Moreover, for these languages where no translation exists researchers can prepare their own version of the PWE (basically any graphic). The same functionality can also be used for creating completely different set of labels for tagging (and also its graphical representation) -- e.g. focused on safety issues or walkability istead of emotion. Although it seems to be very useful feature that increases the flexibility of the system, it should be used very carefully as it also adds some complexity the end-user experience.

Collected data can be accessed and displayed with administration panel separately for each user (see fig. \ref{fig:dashboardsingle}) and for each object (picture). Moreover, data can be exported to CSV file to conduct more sophisticated data analyzes. 

The platform is designed in client-server model and is composed of: mobile application and administration panel hosted on the server and is distributed as an open source\footnote{Please write an email to the authors to get access to the recent version.}.

\subsection{Mobile application}

The mobile application was developed with the use of the Icon 2 framework and dedicated to tables with Android operating system (Android 4.4 or higher). The use of WebView component and well-established framework assures smooth and similar experience on broad variety of tablets (disregard of manufacturer or screen size) and makes it easier to prepare a version for another operating systems (e.g. iOS). Application works only in landscape mode to avoid confusion, which might be caused by two different layouts and switching process. Internet connection is required for application to work, what makes the whole system easier to develop and maintain (no synchronization issues) and at the same time, it is not a problem in urban areas in Poland.  

Each user is limited to only one experiment at a given time. Switching between experiments is possible but it requires logging in into another experiments (login is possible with user name and password or URL link or QR code). This design decision was made on purpose to limit interaction between experiments and make user interface simpler.

\begin{figure}
\centering
\includegraphics[height=105px]{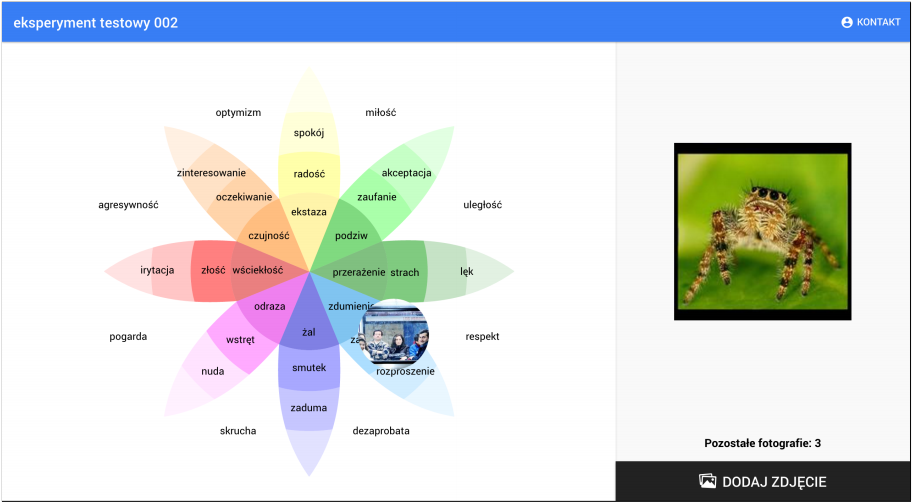}
\caption{The screenshot of the mobile application.}
\label{fig:appfreemode}
\end{figure}

Fig. \ref{fig:appfreemode} shows the screen from the application in the free mode (user makes their own picture). The application window is divided into two parts. On the right appear pictures for evaluation (one-by-one) and on the left, the PWE is displayed. Tagging can be done either by drag-and-drop the picture or by pointing on the wheel (appeared to be important distinction, more about it in the section \ref{sec:testing}). This design is most convenient for right-handed person and has to be inverted for those who are left-handed.

\subsection{Administration panel}

\begin{figure}
\centering
\includegraphics[height=160px]{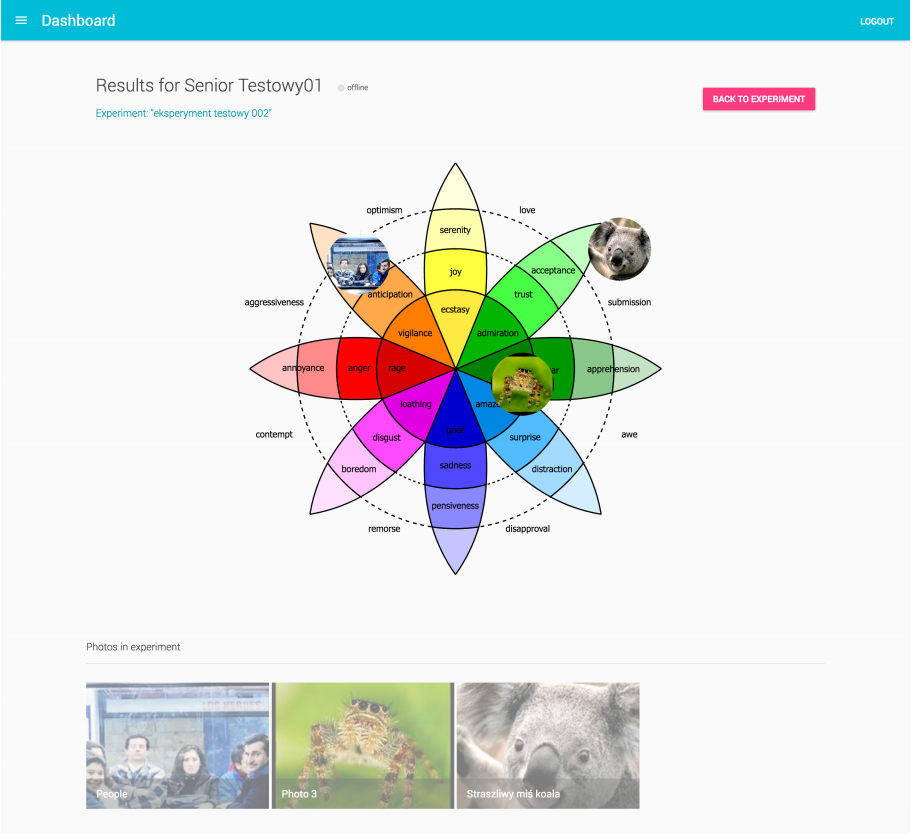}
\caption{Administration panel. Results for a single person. Three pictures placed on the PWE.}
\label{fig:dashboardsingle}
\end{figure}

The administration panel of the platform is based on Phoenix framework and is written in Elixir\footnote{Elixir is a functional language leveraging the Erlang VM and assuring scalability, low-latency and high fault-tolerance.}. Access to the administration panel is possible with any modern web browser. After logging in with user name and password, a researcher can define a new one or edit existing experiment, adding new participants, sending invitations or browsing and exporting results from finished or ongoing experiments (see fig.\ref{fig:dashboardsingle}). 

To start new experiment, a researcher has to decide about: experimental mode (preselected pictures or user's own pictures), start and finish time, participants assigned to this particular experiment and pictures to tag. Almost all decisions can be changed during the experiment (experimental mode is an exception).

The administration panel provides basic views that helps analyzing data (conceived primarily as a tool for quick control of the correctness of a course of an experiment) but for displaying emotions on the map a dedicated GIS systems (or even Google Maps API) has to be used. In a default mode the platform stores exact placing of the picture on the PWE instead of emotion label. For some analysis it might be useful to compare exact placing of different pictures (how far it is from the center). Moreover, it gives a required flexibility when th PWE picture is changed with other tagging map.   

\section{Preliminary tests}
\label{sec:testing}

The test of the platform was precluded by the study of learning and using tables by older adults carried out as a part of PJAIT LivingLab activities\footnote{For more information about LivingLab for older adults check \url{https://livinglab.pja.edu.pl/en/}}. Over 20 older adults (65+) were invited to participate in the location-based game. The scenario of the game was conceived as a combination of fun and learning. The results are thoroughly reported in \cite{kopec2017location} and confirm that using tables is relatively easy to learn even for the elderly who faced it for the first time. Similar results were obtained by Tsai et al.\cite{tsai2015getting}.

The platform (actually the mobile application) was tested with the use of think-aloud protocol on tablet equipped with 10 inch touch screen and Android in version 4.4.2. Four participants, among them two men and two women in age span between 63 and 68, were asked to tag pictures with the mobile application. Users differed considerably in their computer and mobile device skills. The usability study was precluded with a short tutorial of how to use mobile application.

The concept of tagging pictures and labels for emotions was clear and easy to understand for all participants. Nevertheless, few usability issues came out. Firstly, originally implemented mechanism for scaling up pictures needed to be redesigned to be more precise and easier to use. Secondly, experimental mode in which pictures stayed on the left side of the screen after making evaluation was difficult to understand and quickly made the whole screen too crowded. It excludes some usage scenarios as participants will be only able to evaluate each picture separately and reference to previous evaluation will be limited to what they remember and not what they see (the second can be controlled, the first cannot).   

Interesting observations have emerged with respect to the tagging process itself. For more advanced users drag-and-drop mechanism was natural and intuitive, but for less advanced clicking on the target emotion was easier to navigate. The optimal solution would be to choose the method automatically based on user's previous experiences and skills.

Although mobile application can be run not only on tablets, but also on smartphones, the screen size is the limiting factor. The study reveals that 9 -- 10 inch is the minimum screen size for older adults for such application.

\section{Conclusion and future works}

ICTs create opportunities for older people to participate in life and development of cities but dedicated solutions are needed. The platform described in the paper is a step towards including the elderly in the decision making process. The feasibility of the use of the platform by older adults was shown, but further interdisciplinary studies are required to prove that collected data can really be useful for decision making process and urban planning. 

In cooperation with the City of Warsaw and NGOs, two studies are currently planned. The first one will focus on evaluating proposition submitted to the participatory budged for the next fiscal year. The propositions will be translated into pictures and visualization and tagged with help of the platform presented in the paper. Another study is aimed at building an emotion map of Warsaw monuments. Selected group of older adults will be equipped with tables and asked to make and tag pictures of the monuments and commemorative plaques in Warsaw.  

Interviews with older adults during the development of the platform revealed also some hard to overcome limitations (at least within this setup). The elderly limit their outdoor activity when the weather gets worse. Slippery pavements, rain and cold make the task of making pictures of urban spaces really hard. Safety concerns cannot also be ignored as there are many places in the cities of developing countries where taking out  even a cheap tablet on the street is dangerous.

\section{Acknowledgments}
This project has received funding from the European Union's Horizon 2020 research and innovation programme under the Marie Sklodowska-Curie grant agreement No 690962

\bibliographystyle{ACM-Reference-Format}
\bibliography{sigproc} 

\end{document}